\pdfoutput=1

\documentclass[preprint]{sig-alternate-05-2015}

\usepackage[utf8]{inputenc} 

\usepackage{graphicx} 

\usepackage{url}
\usepackage{booktabs} 
\usepackage{array} 
\usepackage{paralist} 
\usepackage{verbatim} 
\usepackage{subfig} 

\usepackage{fancyhdr} 
\usepackage[titles,subfigure]{tocloft} 

\usepackage[T1]{fontenc}
\usepackage[utf8]{inputenc}
\usepackage{authblk}
\usepackage{color}

\title{An analysis of New South Wales electronic vote counting}
\author[ ]{Andrew Conway}
\author[1]{Michelle Blom}
\author[1]{Lee Naish}
\author[1]{Vanessa Teague\thanks{Corresponding author: vjteague@unimelb.edu.au; +61 3 8344 1274.}}
\affil[1]{Department of Computing and Information Systems, University of Melbourne}
\begin{document}
	
\CopyrightYear{2017} 
\setcopyright{acmlicensed}
\conferenceinfo{ACSW '17,}{January 31-February 03, 2017, Geelong, Australia}
\isbn{978-1-4503-4768-6/17/01}\acmPrice{\$15.00}
\doi{http://dx.doi.org/10.1145/3014812.3014837}
\toappear{To appear in ACSW 2017, January 31-February 03, 2017, Geelong, Australia. DOI http://dx.doi.org/10.1145/3014812.3014837}

\maketitle

\begin{abstract}
We re-examine the 2012 local government elections in New South Wales, Australia.  The count was conducted electronically using a randomised form of the Single Transferable Vote (STV).  It was already well known that randomness does make a difference to outcomes in some seats.  We describe how the process could be amended to include a demonstration that the randomness was chosen fairly.  

Second, and more significantly, we found an error in the official counting software, which caused a mistake in the count in the council of Griffith, where candidate Rina Mercuri narrowly missed out on a seat.  We believe the software error incorrectly decreased Mercuri's winning probability to about 10\%---according to our count she should have won with 91\% probability.  

The NSW Electoral Commission (NSWEC) corrected their code when we pointed out the error, and made their own announcement.  

We have since investigated the 2016 local government election (held after correcting the error above) and found two new errors. We notified the NSWEC about these errors a few days after they posted the results.

\end{abstract}

\section{Introduction}
Many Australian elections are tallied electronically, though a computerised count is hard to scrutinise.  Most electoral commissions make  full preference data available, allowing independent recounts, but sometimes not until months after the election.  Some electoral commissions, including Victoria and the Australian Capital Territory, make the counting code openly available online.  Code from the Australian Electoral Commission and  
the New South Wales Electoral Commission (NSWEC) is not available.   

If full source code was available there would be more opportunity to examine the system to find and correct mistakes before, rather than after, the election.   Private software certification is no substitute for public scrutiny.

We reimplemented the NSW count, by reading the NSWEC's recently-released functional specification.  Our re-computation of all the 2012 local government results produced two main results.
\begin{itemize}
\item Randomness significantly impacts some NSW local government outcomes.  This was already well known, but not publicly quantified. Our contribution is a specific suggestion for showing that the randomness is generated fairly.  This is described in Section~\ref{sec:randomness}.    
\item Our count gives a different distribution of preferences in the council of Griffith in 2012.  
The official count contains an error in the computation of the ``last parcel''.  The error decreased Rina Mercuri's winning probability to about 10\% \footnote{private communication verified by a modification of our program} --- according to our count she should have won with roughly 91\% probability. She was not elected. 
\end{itemize}

We are not at all certain that our count is correct, because the specification is ambiguous, the legislation is vague, and of course our code may well contain errors.  
Our source code is openly available for analysis at
{\url{ https://github.com/SiliconEconometrics/PublicService}} and our full results are at {\url{https://siliconeconometrics.github.io/PublicService/} \\ \url{CountVotes/NSWLGE2012MillionRuns/}}.

The rest of this section details the counting process used in  NSW.   We then describe the software error and the effect in Griffith.  Section~\ref{sec:randomness} explains the impact of randomness and how to demonstrate that the random choices are fair.  Section~\ref{sec:ACT} describes software errors that we identified in the ACT code, none of which have affected an election result.  We conclude with a discussion of the implications of our work and the benefits of making election source code available for public scrutiny.

\subsection{The Single Transferable Vote Count}
The Single Transferable Vote (STV) is a proportional, preferential counting method that transforms voter preferences into a set of winning candidates.  

Before counting, we define the Droop quota as

$$Q =  \lfloor v / (s+1) \rfloor + 1 $$

where $s$ is the number of seats to be filled and $v$ is the number of voters.

Counting proceeds in rounds (called ``counts'').  If nobody has a quota, the candidate with the lowest tally is excluded (eliminated) and their votes are redistributed according to the next preference.  (This step is familiar from the UK's ``alternative vote'' and the USA's ``Instant Runoff Vote.'').  Candidates with at least  a quota are elected---their excess over a quota is redistributed according to the next preference.  

This process has a variety of different variants.   For example, what should happen when many candidates attain a quota at the same count?  Should their excesses be distributed immediately, or one after another?  What should happen if a candidate who already has a quota receives votes from another candidate? {\it etc.}

\subsection{New South Wales specifics}
Legislation (from 2005)\footnote{\url{http://www.austlii.edu.au/au/legis/nsw/consol_reg/lgr2005328/sch5.html}} describes the counting process in some detail, but without the clarity of a proper software specification.  Importantly, when a candidate exceeds the quota, the excess votes are distributed using random sampling.  It seems intended that the only votes available for distribution are those received in the last transfer, but no precise definition of last transfer is given.\footnote{However, the more recently written rules for NSW Legislative Council STV counting, which uses the same software, explicitly say, ``Unless all the vacancies have then been filled, the surplus votes of the elected candidate shall be transferred to the continuing candidates in accordance with the provisions of clause 10, but, in the application of those provisions, only those ballot-papers which have been transferred to the elected candidate from the candidate last excluded shall be taken into consideration.'' See {\url{ http://www.legislation.nsw.gov.au/#/view/act/1902/32/sch6}}  Point 14(3).}

The New South Wales Electoral Commission recently published their ``Functional Requirements Specification for the Vote Count,''\footnote{\url{ http://pastvtr.elections.nsw.gov.au/SGE2011/data/Functional_Requirements_for_Vote_Count_v3.2.pdf}}.  This document does add some detail, for example about how to deal with multiple simultaneous elections, but it still omits precise definitions of crucial concepts such as ``count'' and ``last transfer.'' 

\section{The calculation of transferrable votes and the consequence in Griffith}  \label{sec:bug}
The legislation is ambiguous about whether transfers from multiple candidates elected at the same time should count as one transfer when selecting the ``last transfer'' and hence the transferrable votes.  The functional specification deals explicitly with that case, stating in part 1.4.14.1 that:
\begin{quotation}
When Distributing an Elected Candidate’s votes only those votes that have been transferred to a Candidate either from an elected or excluded candidate resulting in that Candidate attaining or exceeding the Quota are taken into consideration. 

These votes may come from more than one transfer of elected candidates if and only if more than one Candidate is elected at a previous Count and Votes from these distributions are transferred to the Candidate resulting in the Quota being attained or exceeded
\end{quotation}

However, pseudocode in the immediately following section (1.4.14.2, 2(c)) begins searching from the \emph{prior} count, looking for the last time a candidate was elected or excluded.  Consider a candidate C who exceeds a quota as a result of the exclusion of candidate E.  Suppose that immediately before E's exclusion some elected candidates' votes were distributed.  In this case, according to 1.4.14.2, 2(c), C's transferrable votes include those elected candidates' votes too.  This is an incorrect computation of the ``last transfer''.  It directly contradicts the quote above:  transferrable votes come from more than one elected candidate, but those distributions did not result in the quota being attained or exceeded.

Exactly this scenario occurred in the council of Griffith in 2012\footnote{Official counts available at {\url{ http://www.pastvtr.elections.nsw.gov.au/LGE2012/Results/LGE2012/PRCC/Griffith/ProgressCount/}}}.

At count 8, Simon Croce is elected.  This results in three surplus transfers in counts 9,10 and~11.
In count 12, Brian Hopper is excluded. As a result, 99 votes go to Paul Rossetto, who thus goes over quota and is elected. (Doug Curran is also elected at this point, but is irrelevant for this discussion.) Paul Rossetto has a greater count than Curran, so Rossetto's surplus votes will be distributed first, and indeed, that is what happens at count 13.

This is shown in the table below:

\begin{center}
\begin{tabular}{clc}
Count & Action & Votes to \\
      &        & Rossetto    \\ 
\hline
 9 & Surplus transfer from  Napoli & 79    \\
10	& Surplus transfer from  Thorpe & 0   \\ 
11	& Surplus transfer from  Croce & 4   \\ 
12	& Exclusion of  Hopper; & 99  \\
    &  Rossetto and  Curran go over quota & \\ 
13	& Surplus transfer from  Rossetto;  & \\
    &  182 prefs distributed & \\
\end{tabular} 
\end{center}

The question is, at count 13, which votes should be transferrable? A reading of 1.4.14.1 seems to imply that the 99 votes transferred to Paul Rossetto in count 12 should be the ones available for redistribution. But what in fact happens in the official count is that 182 votes are available for redistribution. This is a consequence of the logic of 1.4.14.2, which also includes the votes Rossetto got in counts 9,10, and 11, where he got 79, 0 and 4 votes respectively. 

This seems to directly contradict the second paragraph in the 1.4.14.1 quotation above.
83 of the 182 votes do indeed come from more than one transfer of elected candidates, but the votes from these distributions did not result in the quota being reached --- it took the later exclusion of Brian Hopper to do that.
This makes a significant difference as if only 99 votes were available, many of them would be exhausted, and a smaller number of votes could continue at this count. As Alison Balind gets most of these votes, having more at this count makes it easier for her to beat the (otherwise slightly ahead) Rina Mercuri.

Indeed when we run the last transfer calculation as described in 1.4.14.2, our results accord closely with those of the NSW Election Commission, and Alison Balind is elected with very high probability.  However, if we follow 1.4.14.1, and take the transferable votes to be only those received when Brian Hopper was eliminated, the probabilities shift considerably.  Rina Mercury wins with 91\% probability.

We can see how this situation was handled in the prior election to see if it has changed, and indeed it was handled differently in the previous election, clearly done with different software. Consider the 2008 Gwydir general election\footnote{Official counts at \url{ http://vtr.elections.nsw.gov.au/LGE2008/LgeFinalCountReports/Gwydir/Council/GWYDIR - UNDIVIDED.pdf}} where an analogous situation occurs.
In Counts 13 and 14, surpluses were distributed, giving Gordon 7 votes. No one was elected or excluded.
In Count 15, Reardon is excluded, giving Gordon 110 votes to go over quota.
In Count 16, Gordon's excess votes are now distributed, and the last parcel is stated to be 110 votes.
This is shown in the table below:

\vspace{0.2cm}
	\begin{tabular}{clc}
		Count & Action & Votes to  Gordon    \\ 
		\hline
		13 & Surplus transfer from  Egan & 0    \\
		14 & Surplus transfer from  King & 7    \\
		15 & Exclusion of  Reardon & 110 \\ 
		   & Gordon goes over quota &  \\ 
		16	& Surplus transfer from  Gordon; \\
		    &  110 prefs distributed    &  \\
	\end{tabular} 
\vspace{0.2cm}

Using the same algorithm 1.4.14.2 as in the 2012 count, Gordon would have 110+7=117 votes to be distributed. 
So between the 2008 and 2012 election, the vote counting algorithm changed when the software changed.

\subsection{How to correct the 2012 code}
We believe the 2008 logic is correct.  The 2012 logic could be corrected by changing the pseudocode in 1.4.14.2 of the specification, in processing step 2b, by inserting at the start ``If the count when the candidate was elected was an exclusion, set $n=1$. Otherwise...''
The NSW electoral commission informed us that they corrected their program when we pointed out the error.

\section{Randomness in the electronic count}  \label{sec:randomness}

Preference distribution by random sampling means that the same software counting the same
votes may produce a different answer---different margins, different order of election, or even different candidates elected. This was publicised by Anthony Green.
\footnote{\url{http://blogs.abc.net.au/antonygreen/2016/01/} \url{nsw-electoral-law-and-the-problem-of-randomly-elected-} \url{candidates.html} }

We ran our algorithm a million times for each contest; the results are available at {\url{https://siliconeconometrics.github.io/PublicService/CountVotes/NSWLGE2012MillionRuns/}}

For instance, in the Bankstown South Ward, Vanessa Gauci has roughly a one in a million shot of being elected. She was not elected in the official count.
But neither was Anne Connon in the Mosman contest; she had a 96.8\% chance. This is not a result of the discrepancy in the  calculation of ``last parcel''.
When you run a large number of elections with randomised selection for preference distribution, some candidates will lose because of  bad luck. This is a consequence of the legislation, not the implementation.

However, neither the source code for the count nor the method of choosing the randomness are observable by the public.  Without a transparent process showing that the randomness is fairly generated, the outcome could be accidentally or deliberately biased.  There is no evidence of bias, but also no evidence that the random choices were fair. In the next section we show how to provide evidence that the random choices were fair.  Our scheme is easy to implement in time for the next local council elections. 

\subsection{Demonstrating fair randomness}

Elections generally aim to be fair and to be seen to be fair. The legislated randomness makes life difficult for the NSWEC, who have to implement
a randomized process, and be seen to be fair and transparent about it. It is desirable for the election commission to be able to defend itself against
a candidate who had bad luck and asserts that the count was biased; it would also be desirable to defend against a candidate who demands
a recount hoping for better luck.

Surprisingly, this is entirely consistent with an electronic count, provided some simple conditions are met. There is a well known concept of pseudo-random number generators. This is a 
computational device that, given a starting number (called a seed), produces a long series of numbers that have many of the properties of random numbers, indeed enough to fairly
implement the NSW count. If you start with the same
seed, you get the same series. This is how computers generally produce the semblance of randomness. 

If you make this seed public, the details of the pseudo-random number generator public, and the source code of the counting  software public, then anyone else can in principle
replicate the ``random'' choices exactly to see that they are done fairly.

Of course, choosing the seed is important.  
After the full preference data file has been published, including at the counting ceremony, the NSWEC could have a public
process for generating the random seed---for example, using dice or a machine like the Tattslotto machine.
Philip Stark's tools for Risk Limiting Audits\footnote{\url{https://www.stat.berkeley.edu/~stark/Java/Html/sha256Rand.htm}}  combine a transparent process for initialising the randomness (such as by throwing dice) with a publicly verifiable process for transforming that randomness, 
using a pseudo-random number generator, into a long list of random choices.  This technique could carry over immediately into the NSW count.  

As well as demonstrating the fairness of the count, this would have the side benefit of making recounts produce exactly the same result assuming
no errors are found in the entry of the paper ballots.

\section{Analysis of the ACT vote counting program}
\label{sec:ACT}
The ACT electoral commission does make their vote counting program available.
The Logic and Computation Group at the Australian National University have found three bugs in the vote-counting module of eVACS to date.

The first bug was a simple for-loop bound error. The code would work correctly or fail depending on whether the number of candidates was even or odd. They found it days before the system was going to be used in a live televised election count. ACT Elections acknowledged the bug but asserted that they would have found the bug immediately upon starting the live count.

The second bug was actually an error in the legislation. To break multi-way ties for a single weakest candidate, the Hare-Clarke method compares the tallies of the tied candidates in previous rounds. The legislation states that one should go back to the previous round in which all candidates have an unequal number of votes. So if there are three tied weakest candidates A, B and C, then one has to return to the previous round in which their tallies are {\em pairwise distinct}. In the worst case, one may have to return all the way to the first count in which all their tallies are 0. When they reported this bug, ACT Elections confirmed that they knew about the bug and that eVACS used a more sensible approach where they return to the previous round where one of the candidates is weaker than the others. They showed that there are elections where different choices for breaking such ties can lead to different results. 

The third bug was an initialisation error where the code declared a boolean flag but did not initialise it at the start. They found that different C compilers gave different results since they initialise this flag in opposite ways. They also found example elections where this difference could lead to different results. The bug was acknowledged by ACT Elections and repaired.

They have also found two errors in the ACT Elections Fact Sheet which outlines the counting procedure in plain language. ACT Elections has acknowledged these bugs and have also noted that they have been present and have gone unnoticed for 15 years.

This is a success story for the ACT electoral commission. Some bugs were fixed before they became an issue.

The NSW legislation and specification has an almost identical issue for three way tie resolution (sections 1.4.8.1 and 1.4.26 in the specification). 
We implemented a reasonable interpretation similar to the ACT's resolution;
the NSW election commission implemented an alternative reasonable interpretation (private communication) but have not at the time of writing incorporated it into the specification
or published it with the specification. This is an exceedingly rare issue but should on principle be clarified.

\section{Implication of bugs in certified election software}
\label{sec:implications}
Software is notorious for being buggy. Humans can build vastly complex software projects much faster than any other type of engineering. Humans are poor at understanding the complexity as a whole, and testing it, and therefore
making bug free computer programs is exceedingly difficult. In some circumstances formal verification is possible, but this is exceedingly difficult. 

The NSWEC tested and certified their software. An expert in vote counting certified the specification as representing the law.\footnote{\url{https://www.elections.nsw.gov.au/__data/assets/pdf_file/0003/171651/PRCC_Fn_Spec_v3.2_Certificate_of_Legislative_Compliance_-_Final.pdf}} Then an Indian testing company, Birlasoft,
certified the software as representing the specification.\footnote{\url{https://www.elections.nsw.gov.au/__data/assets/pdf_file/0004/171652/PRCC_LG_Birlasoft_Test_Certificate_v3.2.pdf}} 
None of them noticed the algorithmic error or the inconsistency in the specification. This is not unreasonable --- that section of the specification is quite hard to interpret. 
We only noticed the problem after spending days trying to reconcile our results with the NSWEC's results. 

The important lesson from this is that certification of election software is not a reason to trust it.

It means that subtle errors may not be noticed for years, if at all. 
It is fortunate that this one is observable; the counts that the NSW election commission put on the web provide enough information to detect the issue, even if it took years.
The NSWEC did not release the raw votes for the 2015 state election until after the deadline for disputing returns had passed, months after the election,
too late for anyone to actually check it.

Of a much more serious concern is the invisible software that produces the file containing the list of votes to be counted.  An error or security vulnerability in that code might change an election result without there being any way for an external observer to detect the problem.  This too is easy to address: there should be an opportunity for scrutineers to audit the paper ballots against the published electronic full preference data file.

Since some votes were cast electronically in the 2015 NSW election using the iVote Internet voting platform 
in a way that made it impossible for any scrutineers to verify, it is impossible for anyone
to verify that the list of votes from iVote is correct. Of course, a malicious hacker who changed the results could know that the file was {\it incorrect}.

\section{Public access to election software source code}

There are many reasons why the electoral commissions in Australia should make their code public before the election.
\begin{itemize}
	\item It enables external people to notice bugs {\em before} the software is used in an election.
	\item It makes it easier for external people to verify bugs. Our software produced a different result --- probabilistically --- to the official count. If we had 
	not had access to some of NSWEC's probabilistic experimentation (private communication from NSWEC), the error in Griffith would have been more difficult to find.
	\item It is necessary to demonstrate fair randomness for the NSW randomised counting algorithm.
	\item It makes it easier for the electoral commission to demonstrate and defend its integrity.
\end{itemize}

This is not just a NSW issue.
There has been an effort to obtain access to the Australian Electoral Commission's counting program under freedom of information laws.
There was also a senate motion requesting publication of the code\footnote{\url{http://lee-rhiannon.greensmps.org.au/content/news-} \url{stories/update-public-release-secret-senate-voting-system}}.
These were resisted 
vigorously and successfully by the Australian Electoral Commission on the basis of security concerns and commercial value of the program. The general
consensus amongst security experts is that programs subject to public scrutiny are more secure.  Openness does not guarantee that all
errors or security problems will be detected, but neither does private certification. The commercial argument seems a somewhat
weak argument since (from practical experience) we found that writing a counting program is only a couple of days' work. Testing it adequately of course takes longer, but public scrutiny would help this.

\section{2016 NSW local government election}

The NSWEC acknowledged and corrected that bug before the 2016 local government election. However they did not make source available nor
the randomness source public.

We checked the results from the 2016 election, and
found two new errors that resulted in incorrect distribution of preferences in four instances. 
Our guess is that they did not affect who was elected, but we are unable to confirm this without access to the source code. We reported these errors to the NSWEC within a week of the
results being posted (within the time period for challenges), but have not had any meaningful response from them. 

\subsection{Three way ties for elimination}

The general approach for STV, when a point in the tally is reached where all votes above a quota have been redistributed, but not enough candidates have been elected,
is for the candidate with the lowest tally to be chosen for exclusion. Each paper assigned to that candidate is distributed to the next
continuing candidate on the paper's preference list. Specific instances of STV have assorted technicalities around this; one of these is what to
do in the case of ties for the minimum tally. 

The NSW specification (section 1.4.26) says that for local government elections, the tie should be resolved by looking at
prior progressive totals (a {\em countback}), and excluding the candidate who most recently had a lowest tally. If no such tie break occurs, a random draw is performed. 
The specification is somewhat ambiguous in multi-way ties, but a suitable ruling has been established and given to us informally.

In the Hawkesbury contest however, count 9, Shaun Middlebrook was incorrectly excluded rather than Michelle Carter, who had a lower count in the countback (on count 2).
Similarly in count 20, Jay Graeme was excluded incorrectly instead of John Thomas, who was lower on count 11.

In Campbelltown, count 17, Youssed Raid was incorrectly excluded instead of Carla Simmons, who was lower on count 14. 

In each of these three instances, there was a three way tie. Possibly that was not tested.

This probable bug happened to not make a difference in this case, as Campbelltown’s results are the same when run a million times without this putative bug. 
Hawkesbury is similar, although it may change the order of election. Not having access to the source code we cannot confirm that the bug is restricted to these two contests.

\subsection{Rounding of votes to be transferred during excess distribution}

When distributing the excess from an elected candidate, in NSW the choice of excess is implemented defined by transferring only a portion (defined by a transfer value between 0 and 1) of the votes.
This is done by distributing the appropriate papers to each candidate based on the next continuing preference. For each candidate, the number of papers distributed is
multiplied by the transfer value to get the number of papers transferred. Usually, this number will not be an integer. 
Rounding is done by a set of rounding rules defined in section 1.4.17.1 of the specification.
A fixed number will be rounded up, ordered by largest fractional value (as makes sense). In case of ties, preference goes to the largest integer value, then highest values in a countback, then a random draw.
In Bland Shire Council, count 2, the transfer value is exactly 0.2. Four candidates end up with a fractional value of 0.6; three of those 4 need to be chosen. The integers are:

\begin{tabular}{ll}
MONAGHAN Brian & 9 \\
GRELLMAN Peter & 7 \\
BAKER Bruce & 5 \\
THOMAS Muzz & 3 \\ 
\end{tabular}

It is unambiguous that the first three should be rounded up, and the last rounded down. But instead, Bruce Baker is the candidate who was rounded down, and Muzz Thomas was rounded up.
There is some rounding that the specification requires to fixed number of decimal digits. However, this does not affect the outcome as decimal rounding does not affect multiples of 0.2. 

In the specific case of Bland Shire, the same candidates are elected in a million runs (although it may affect order of election).
The more serious effect is what this could have on random number generation (in case the bug causes a random draw, which is somewhat unlikely to cause the observed outcome, but other possibilities are also somewhat problematic). This could have a large knock on effect. In particular, a very similar situation arises in count 3 of Yass Valley Council. In this particular case the NSWEC rounding appears correct, but an undetectable (from the outside) effect on random number generation is likely to change the result of the election, as our million runs show Greg Butler should be elected roughly 56\% of the time (but wasn’t in the official count) while Kim Turner should be elected roughly 36\% of the time (and was in the official count). Not having access to the source code we cannot confirm that this result was not affected by this bug.

\section{Conclusion}

Randomness in the NSW counting legislation makes it challenging to demonstrate the fairness of the count.  This can be resolved by making software source code public, and having a public ceremony for generating the random seed after the full preference data file is published.

The code for the NSW local government count was incorrect, despite certification. This probably impacted the election 
outcome in Griffith in 2012. The error found was corrected, but in the next election we detected another two apparent errors.
 We could detect this only because we could verify the count directly and find a mistake. 

There are other computerized systems critical for the election in NSW and elsewhere that cannot be verified by the public because the inputs are not available.

When these processes were conducted on paper, scrutineers insisted on observing until they were confident the proper process had been followed.  When computers are involved, the same scrutiny is necessary, for the same reasons.  

It would be good for democracy, and good for the electoral commissions, to make election-related source code public before an election.  That doesn’t guarantee that the software is correct or secure, but it raises the likelihood that errors will be identified and corrected before an election. 
\end{document}